\documentclass[conference]{IEEEtran}

\IEEEoverridecommandlockouts

\usepackage[utf8]{inputenc}
\usepackage[T1]{fontenc}
\usepackage{graphicx}
\usepackage{dblfloatfix}
\usepackage{float}
\usepackage{filecontents}
\usepackage[noadjust]{cite}
\usepackage[hidelinks]{hyperref}
\usepackage{balance}
\usepackage{flushend}
\usepackage{amsmath,amssymb,amsfonts}
\usepackage{algorithmic}
\usepackage{textcomp}
\usepackage{xcolor}
\usepackage{mdframed}
\usepackage{enumitem}
\usepackage[american]{babel}
\usepackage{subcaption}
\usepackage{multirow}
\usepackage{csquotes}
\usepackage{listings}
\usepackage[skins]{tcolorbox}

\usepackage{fancyhdr}

\newtcbox\tcbtp{hbox, on line, colback=lightgray, enhanced, frame hidden, boxrule=-1pt, 
    top=-2pt, bottom=-2pt, right=-1pt, left=-1pt}

\definecolor{link}{rgb}{0.63, 0.79, 0.95}

\def\BibTeX{{\rm B\kern-.05em{\sc i\kern-.025em b}\kern-.08em
    T\kern-.1667em\lower.7ex\hbox{E}\kern-.125emX}}

\makeatletter
\patchcmd{\@maketitle}
  {\addvspace{0.5\baselineskip}\egroup}
  {\addvspace{-0.5\baselineskip}\egroup}
  {}
  {}
\makeatother

\begin{document}

\title{\huge KG-EmpiRE: A Community-Maintainable Knowledge Graph for a Sustainable Literature Review on the State and Evolution of Empirical Research in Requirements Engineering\vspace{-0.2cm}}



\author{\IEEEauthorblockN{Oliver Karras\IEEEauthorrefmark{1}}
\IEEEauthorblockA{\IEEEauthorrefmark{1}TIB - Leibniz Information Centre for Science and Technology, Germany, Email: oliver.karras@tib.eu}}



\definecolor{link}{RGB}{0, 123, 255}
	


\maketitle


\begin{abstract}
In the last two decades, several researchers provided snapshots of the ``current'' state and evolution of empirical research in requirements engineering (RE) through literature reviews. However, these literature reviews were not sustainable, as none built on or updated previous works due to the unavailability of the extracted and analyzed data. KG-EmpiRE is a Knowledge Graph (KG) of empirical research in RE based on scientific data extracted from currently 680 papers published in the IEEE International Requirements Engineering Conference (1994-2022). KG-EmpiRE is maintained in the Open Research Knowledge Graph (ORKG), making all data openly and long-term available according to the FAIR data principles.
Our long-term goal is to constantly maintain KG-EmpiRE with the research community to synthesize a comprehensive, up-to-date, and long-term available overview of the state and evolution of empirical research in RE.
Besides KG-EmpiRE, we provide its analysis with all supplementary materials in a repository. This repository contains all files with instructions for replicating and (re-)using the analysis locally or via executable environments and for repeating the research approach.
Since its first release based on 199 papers (2014-2022), KG-EmpiRE and its analysis have been updated twice, currently covering over 650 papers. 
KG-EmpiRE and its analysis demonstrate how innovative infrastructures, such as the ORKG, can be leveraged to make data from literature reviews FAIR, openly available, and maintainable for the research community in the long term. In this way, we can enable replicable, \mbox{(re-)usable}, and thus sustainable literature reviews to ensure the quality, reliability, and timeliness of their research results.
\end{abstract}

\begin{IEEEkeywords}
Knowledge graph, empirical research, requirements engineering, FAIR, sustainability, literature review
\end{IEEEkeywords}



\section{Introduction}
For 20 years, various researchers conducted literature reviews to examine the state and evolution of empirical research in requirements engineering (RE) with the shared goal of providing a comprehensive, up-to-date, and long-term available overview~\cite{Karras.2023b, Karras.2023}. However, these literature reviews were not sustainable, as none built on or updated previous ones, which are known challenges of literature reviews~\cite{dos_Santos.2021}. While recent research addresses these challenges by providing social and economic decision support and guidance~\cite{dos_Santos.2021}, the underlying problem is the unavailability of the extracted and analyzed data. Researchers need technical support, i.e., infrastructures, to conduct sustainable literature reviews so that all data is openly and long-term available according to the FAIR data principles~\cite{dos_Santos.2021} and corresponding to open science in SE~\cite{Mendez.2020}.

In their joint work, Wernlein~\cite{Wernlein.2022} and Karras et al.~\cite{Karras.2023b, Karras.2024} examined the use of the Open Research Knowledge Graph (ORKG)~\cite{Stocker.2023}, as such technical support by building, publishing, and analyzing a Knowledge Graph (KG) of empirical research in RE \mbox{(KG-EmpiRE)} based on \textit{currently} 680 research track papers of the IEEE International Requirements Engineering Conference (1994-2022).\vspace{0.2cm}

In this paper, we present the KG-EmpiRE, available in the ORKG\footnote{\url{https://orkg.org/observatory/Empirical_Software_Engineering}}, and its analysis, available on GitHub~\cite{Karras.2023c}, Zenodo~\cite{Karras.2024a}, and on Binder\footnote{\url{https://tinyurl.com/empire-analysis}} for interactive replication and (re-)use.

KG-EmpiRE contains scientific data on the six themes~\textit{research paradigm}, \textit{research design}, \textit{research method}, \textit{data collection}, \textit{data analysis}, and \textit{bibliographic metadata}. We plan to expand these themes in the long term. For more details on the themes, refer to the supplementary materials~\cite{Karras.2024a, Karras.2023c}. Since its first release based on 199 papers (2014-2022)~\cite{Wernlein.2022}, KG-EmpiRE and its analysis have been updated twice. \mbox{Karras~et~al.}~\cite{Karras.2023b} published the first update with 570 papers (2000-2022) at the 17th ACM/IEEE International Symposium on Empirical Software Engineering and Measurement 2023, where they received the best paper award. The second update is ongoing and covers 680 papers (1994-2022) so far. The goal for the second update is to cover all 748 research track papers from the IEEE International Requirements Engineering Conference~\mbox{(1993-2023)}. 

The analysis provides answers to 16 out of 77 competency questions (cf. supplementary materials~\cite{Karras.2024a, Karras.2023c}) regarding empirical research in RE that we derived from the vision of Sjøberg et al.~\cite{Sjoberg.2007} on the role of empirical methods in SE, including RE, for 2020-2025. While the number of competency questions answered reflects the coverage of the curated topic in KG-EmpiRE, the answers to competency questions provide insights into the state and evolution of empirical research in RE. For each competency question answered, we provide all details of the analysis with its data, visualizations, explanations, and answers in a repository~\cite{Karras.2024a, Karras.2023c} that is also hosted on Binder for interactive replication and (re-)use.

Overall, this repository contains all files with detailed explanations and instructions for replication and (re-)use of \mbox{KG-EmpiRE} and its analysis locally or via executable environments (Binder and GitHub Codespaces), as well as for repeating the research approach for sustainable literature reviews with the ORKG. The repository also contains all generated visualizations with their data, exported as PNG and CSV files, as well as supplementary materials on the themes, their structuring in the ORKG, and all 77 competency questions.

\section{Structure of KG-EmpiRE and the Repository}

\subsection{KG-EmpiRE}
We developed an ORKG template\footnote{\url{https://orkg.org/template/R186491}} to organize the scientific data extracted from the papers in the ORKG. ORKG templates implement a subset of the Shapes Constraint Language (SHACL) and allow specifying the underlying (graph) structure to organize the data in a structured manner~\cite{Hussein.2023}. In this way, we determined which data to extract and standardized their description to ensure they are FAIR, consistent, and comparable across all papers. The developed ORKG template covers the six themes investigated. For more details on the ORKG template, refer to the supplementary materials~\cite{Karras.2024a, Karras.2023c}.

By applying the ORKG template to the papers, KG-EmpiRE currently consists of almost 35,000 triples, which are made up of over 51,000 resources and almost 19,000 literals (see \autoref{tab:stats}). While these statistics reflect the efforts to provide a solid structured description of the extracted data, they also show that KG-EmpiRE is relatively small compared to the entire ORKG and other well-known knowledge graphs, e.g., Wikidata or DBpedia, which include millions of entities.

\begin{table}[htbp]
\centering
\caption{Statistics of KG-EmpiRE and the ORKG on 30.04.24.}
\label{tab:stats}
\begin{tabular}{|c|c|c|c|c|}
\hline
\textbf{Type} & \multicolumn{1}{l|}{Paper} & \multicolumn{1}{l|}{Triples} & \multicolumn{1}{l|}{Resources} & \multicolumn{1}{l|}{Literals} \\ \hline \hline
\textbf{KG-EmpiRE} & 680 & 34,912 & 51,035 & 18,789 \\ \hline
\textbf{ORKG} & 29,379 & 1,407,164 & 432,323 & 768,212 \\ \hline
\end{tabular}
\vspace{-0.2cm}
\end{table}

\subsection{Repository}
In the repository, there are three folders and six files, with the Jupyter Notebook \textit{empire-analysis.ipynb} as the main file. The Jupyter Notebook encapsulates the entire analysis of KG-EmpiRE and provides visualizations, explanations, and answers for each of the 16 competency questions. The visualizations are exported as PNG files per competency question to the \textit{Figures} folder. The data retrieved by KG-EmpiRE for analysis is stored as CSV files for each competency question in the \textit{SPARQL-Data} folder by date. In this folder, we also provide CSV files of the latest release to replicate the results of the related publication~\cite{Karras.2023b}. The last folder \textit{Supplementary materials} provides additional materials for detailed overviews of the content for data extraction regarding the themes, the developed ORKG template, all 77 competency questions derived, and the research approach. The second most important file is \textit{README.md}, which contains detailed explanations and instructions about the project, the repository, its installation (locally and via executable environments), the replication of the analysis, and the (re)use of KG-EmpiRE with its most recent data. The remaining four files support the installation (\textit{requirements.txt}, \textit{runtime.txt}), clarify the copyright (\textit{LICENSE}), and ensure the citability of the repository (\textit{CITATION.cff})\footnote{\url{https://citation-file-format.github.io/}}.

\section{Conclusion}
Overall, KG-EmpiRE and its analysis lay the foundation for a sustainable literature review on the state and evolution of empirical research in requirements engineering. They can be used to replicate the results from the related publication~\cite{Karras.2023b}, (re-)use the data for further studies, and repeat the research approach for sustainable literature reviews on other topics. KG-EmpiRE and its analysis demonstrate how innovative infrastructures, such as the ORKG, can be leveraged to make data from literature reviews FAIR and openly available in the long term. In this way, researchers can build on and update the data ideally collaboratively, enabling sustainable literature reviews for comprehensive, up-to-date, and long-term available overviews, true to the principle: \textit{Divide et Impera}.

In summary, the special feature of KG-EmpiRE lies in the proof that data from literature reviews can already be prepared during data extraction in such a way that they are understandable and processable by humans and machines to update, replicate, and (re-)use them sustainably. KG-EmpiRE and the underlying research approach using technical infrastructures, such as the ORKG, have the potential to be used on a large scale to establish sustainable literature reviews and thus ensure the quality, reliability, and timeliness of their research results.


\section*{Acknowledgment}
\footnotesize The authors thank the Federal Government, the Heads of Government of the Länder, as well as the Joint Science Conference (GWK), for their funding and support within the NFDI4Ing and NFDI4DataScience consortia. This work was funded by the German Research Foundation (DFG) project numbers 44214671 and 460234259 and by the European Research Council for the project ScienceGRAPH (Grant agreement ID: 819536).

\bibliographystyle{IEEEtran}
\balance
\bibliography{IEEEabrv,references}

\begin{thebibliography}{10}
\providecommand{\url}[1]{#1}
\csname url@samestyle\endcsname
\providecommand{\newblock}{\relax}
\providecommand{\bibinfo}[2]{#2}
\providecommand{\BIBentrySTDinterwordspacing}{\spaceskip=0pt\relax}
\providecommand{\BIBentryALTinterwordstretchfactor}{4}
\providecommand{\BIBentryALTinterwordspacing}{\spaceskip=\fontdimen2\font plus
\BIBentryALTinterwordstretchfactor\fontdimen3\font minus
  \fontdimen4\font\relax}
\providecommand{\BIBforeignlanguage}[2]{{%
\expandafter\ifx\csname l@#1\endcsname\relax
\typeout{** WARNING: IEEEtran.bst: No hyphenation pattern has been}%
\typeout{** loaded for the language `#1'. Using the pattern for}%
\typeout{** the default language instead.}%
\else
\language=\csname l@#1\endcsname
\fi
#2}}
\providecommand{\BIBdecl}{\relax}
\BIBdecl

\bibitem{Karras.2023b}
O.~Karras, F.~Wernlein, J.~Kl{\"u}nder, and S.~Auer, ``{Divide and Conquer the
  EmpiRE: A Community-Maintainable Knowledge Graph of Empirical Research in
  Requirements Engineering},'' in \emph{International Symposium on Empirical
  Software Engineering and Measurement}, 2023.

\bibitem{Karras.2023}
\BIBentryALTinterwordspacing
O.~Karras, F.~Wernlein, J.~Klünder, and S.~Auer, ``\BIBforeignlanguage{en}{{A
  Comparison of Scientific Publications on the State of Empirical Research in
  Requirements Engineering and Software Engineering}},'' 2023. [Online].
  Available: \url{https://orkg.org/comparison/R650023/}
\BIBentrySTDinterwordspacing

\bibitem{dos_Santos.2021}
V.~{dos Santos} \emph{et~al.}, ``{Towards Sustainability of Systematic
  Literature Reviews},'' in \emph{15th ACM/IEEE International Symposium on
  Empirical Software Engineering and Measurement}, 2021.

\bibitem{Mendez.2020}
D.~Mendez \emph{et~al.}, ``{Open Science in Software Engineering},'' in
  \emph{Contemporary Empirical Methods in Software Engineering}.\hskip 1em plus
  0.5em minus 0.4em\relax Springer, 2020.

\bibitem{Wernlein.2022}
\BIBentryALTinterwordspacing
F.~Wernlein, ``{Acquisition and Analysis of Research Practices Using Semantic
  Structures},'' {Bachelor Thesis}, Gottfried Wilhelm Leibniz Universit{\"a}t,
  2022. [Online]. Available: \url{https://doi.org/10.15488/12796}
\BIBentrySTDinterwordspacing

\bibitem{Karras.2024}
O.~Karras, F.~Wernlein, J.~Kl{\"u}nder, and S.~Auer, ``{KG-EmpiRE: A
  Community-Maintainable Knowledge Graph of Empirical Research in Requirements
  Engineering},'' in \emph{Software Engineering 2024}, 2024.

\bibitem{Stocker.2023}
M.~Stocker \emph{et~al.}, ``{FAIR Scientific Information with the Open Research
  Knowledge Graph},'' \emph{FAIR Connect}, vol.~1, no.~1, 2023.

\bibitem{Karras.2023c}
\BIBentryALTinterwordspacing
O.~Karras, ``{Divide and Conquer the EmpiRE: A Community-Maintainable Knowledge
  Graph of Empirical Research in Requirements Engineering - A Sustainable
  Literature Review for Analyzing the State and Evolution of Empirical Research
  in Requirements Engineering},'' 2023. [Online]. Available:
  \url{https://github.com/okarras/EmpiRE-Analysis}
\BIBentrySTDinterwordspacing

\bibitem{Karras.2024a}
\BIBentryALTinterwordspacing
------, ``{Divide and Conquer the EmpiRE: A Community-Maintainable Knowledge
  Graph of Empirical Research in Requirements Engineering - A Sustainable
  Literature Review for Analyzing the State and Evolution of Empirical Research
  in Requirements Engineering},'' 2024. [Online]. Available:
  \url{https://doi.org/10.5281/zenodo.11092471}
\BIBentrySTDinterwordspacing

\bibitem{Sjoberg.2007}
D.~I. Sjoberg \emph{et~al.}, ``{The Future of Empirical Methods in Software
  Engineering Research},'' in \emph{Future of Software Engineering}, 2007.

\bibitem{Hussein.2023}
H.~Hussein \emph{et~al.}, ``{Increasing Reproducibility in Science by
  Interlinking Semantic Artifact Descriptions in a Knowledge Graph},'' in
  \emph{Leveraging Generative Intelligence in Digital Libraries: Towards
  Human-Machine Collaboration}.\hskip 1em plus 0.5em minus 0.4em\relax
  Springer, 2023.

\end{thebibliography}

\end{document}